\documentclass{aa}
\usepackage{graphicx}
\usepackage{natbib}
\usepackage{amssymb}
\usepackage{subfigure}
\bibpunct{(}{)}{;}{a}{}{,} % to follow the A&A style

\begin{document}
\titlerunning{Star formation in the swept-up shells of the Rosette Nebula}
\authorrunning{Li \& Smith}
\title{Multi-seeded multi-mode formation of embedded clusters in the RMC: 
Clusters formed in swept-up shells
\thanks{This publication makes use of 2MASS, the Two Micron All Sky Survey, a
joint project of the University of Massachusetts and the Infrared Processing
and Analysis Center, funded by the National Aeronautics and Space administration
and the National Science Foundation  } 
     }
%\subtitle{Clusters baked in swept-up shells }
  
\author{J. Z. Li\inst{1,2} \& M. D. Smith\inst{1}}  

\institute{National Astronomical Observatories, Chinese Academy of 
           Sciences, Beijing 100012, China\\
           \email{ljz@ns.bao.ac.cn}
      \and Armagh Observatory, College Hill, Armagh BT61 9DG, N. Ireland, UK
} 

\date{Received ...; / Accepted ... }
 \author{J. Z. Li\inst{1,2} \& M. D. Smith\inst{1}}
\offprints{J. Z. Li: \email{ljz@bao.ac.cn}}

\abstract{This is the first of a series of three papers on clustered
star formation in the Rosette Molecular Complex. Here we investigate star formation in 
the interfacing layers between the 
expanding Rosette Nebula and its surrounding cloud, based on an analysis of the spatially complete and 
unbiased 2MASS data. Two medium-mass infrared clusters with ages of around 1~Myr are 
identified in the south and south-east arcs of the fragmented shell. The majority of the 
candidate cluster members in these radiation and pressure-confined regions are found to be 
almost uniformly distributed, roughly following the compression layers traced by the
distribution of optical depth at 100 $\mu$m, and may well develop into gravitationally 
unbound systems upon their emergence from the parental cloud. These expanding 
shells are believed to be playing important roles in impeding the emerging young
open cluster NGC~2244 from intruding immediately and deeply into the ambient
molecular cloud, where sequential formation of massive clusters is taking place.
%on time scales of a few Myrs.

\keywords{ISM: clouds -- Infrared: stars -- Stars: formation -- Stars: pre-main sequence}
}
\maketitle

%%%%%%%%%%%%%%%%%%%%%%%%%%%%%%%%%%%%%%%%%%%%%%%%%%%%%%%%%%%%%%%%%%%%%%%%%%%%%%%%
\section{Introduction}
%%%%%%%%%%%%%%%%%%%%%%%%%%%%%%%%%%%%%%%%%%%%%%%%%%%%%%%%%%%%%%%%%%%%%%%%%%%%%%%%

Most stars, if not all, are believed to originate as members of either groups
or clusters \citep{1991ApJ...371..171L}. The early clustering has been 
partly catalogued through searches at infrared wavelengths,
which are able to locate the stars even when they are still 
deeply embedded in molecular clouds \citep{2003AJ....126.1916P,2003A&A...397..177B}.
Since the advent of the 2 Micron All Sky Survey (2MASS) large samples of infrared clusters have 
been discovered toward the Galactic plane \citep{2003A&A...404..223B,2003A&A...400..533D},
projected over the Galactic center \citep{2000A&A...359L...9D} as well as in
well-known molecular clouds at high Galactic latitudes \citep{2000AJ....120.3139C}.

In order to trace the many possible
evolutionary paths from molecular clouds to either field stars or bound systems, 
it is vital to select and investigate specific samples of embedded clusters.
Such a study, undertaken here for the Rosette Molecular Complex, can provide insight 
into (1) the mechanisms responsible for the fragmentation of giant molecular clouds (GMCs), 
(2) the conditions essential for the development of clumpy cores into stellar clusters, 
(3) the dominant modes of star and cluster formation, (4) the dynamical relaxation of
clusters and, ultimately, (5) the star formation history of the Galaxy.

The Rosette Molecular Cloud (RMC) is an ideal location for testing out scenarios for 
massive star formation on large spatial scales. At a distance of 1.4~kpc
\citep{2000A&A...358..553H}, it is of mass 10$^5$~M$_\odot$ and extends over 100\,pc with a 
width of up to 50\,pc, displaying a striking interaction with the spectacular Rosette Nebula 
towards one end \citep{1980ApJ...241..676B}. The latter H{\small II} region has a radius of
15\,pc (38\arcmin). It is encompassed by a shell of H{\small I} of radius 20\,pc (48\arcmin) 
which is expanding into the cloud at 4.5~km~s$^{-1}$ \citep{1993ApJ...414..664K}. 
The neutral shell is also associated with molecular clumps and strong 100$\mu$m emission 
\citep{1995ApJ...451..252W}. The subject of this work is the embedded stars in this shell.

The gas is clumped on all measured scales down to 0.1~pc and 1~M$_\odot$ 
\citep{1995ApJ...451..252W,1998A&A...335.1049S}. The clump number-mass spectral index is 
$\alpha \sim 1.3 - 1.6$ (depending on how sub-clumps are treated) as fitted by the distribution  
dN/dM $\propto$ M$^{-\alpha}$. In space, the clumps are distributed around a ridge running 
parallel to, but two degrees below, the Galactic plane \citep{1995ApJ...451..252W}. 

Many high mass stars, especially those immersed in the Rosette Nebula, interact with the cloud. 
Near-infrared observations have already
revealed several prominent clusters on the scale of 1 parsec, as listed by 
\citet{1997ApJ...477..176P}. The cluster sizes and their absence on optical images indicate that they
are young and embedded. \citet{1997ApJ...477..176P} suggested that triggered star formation 
is rife but not an exclusive mechanism, with spontaneous formation as an alternative. 
It should be noted that the rate for embedded clusters to emerge as bound clusters is 
statistically low in the Galaxy, based on the assumption of continuous star formation 
and if compared to known young open clusters \citep{2003ARA&A..41...57L}. 
Therefore, we should enquire whether the RMC clusters are in the process of dispersing. 
The RMC clusters, however, may be too young to display signs of dispersal given
that the nebula, powered by the cluster NGC\,2244 with a main-sequence turn-off
 age of 2\,Myr \citep{2002AJ....123..892P}, possesses a dynamical age of just 0.2--0.6\,Myr
\citep{1967ApJ...147..965M}. In comparison, however, the dynamic age of the H{\small I} shell 
is 4~Myr. If we take a stellar velocity dispersion of 2~km~s$^{-1}$, then 
an unbound system would expand by 2\,pc (5\arcmin) in 1\,Myr. Hence, some dissolution may be
expected on this scale.

The RMC was reported by \citet{2004LSa} to be pervaded by new-born clusters displaying a
multi-seeded origin at various sites. 
%Embedded clusters from medium to high-mass appear 
%either as compact or loose aggregates, widely distributed but in a neatly structured manner 
%over the south-east quadrant of the molecular complex.  
Following perhaps the emergence of 
the young OB cluster NGC 2244, both triggered and non-triggered cluster formation are 
in evidence along the major axis of the complex.
It is taking place in or probably adjacent 
to the swept-up layers of the HII region and deep in the cloud toward its south-east
boundary, respectively.
A detailed study of each spatially confined region has been carried out to substantiate the
embedded nature of the clusters. Here, in the first of this series
of papers, we focus on the cluster formation in the fragmented interaction layer of the
HII region with the molecular cloud.
%We ask: how do the distributions in space and mass compare to that of the initial cores
%and the final stars?
Massive compact subclusters were identified in the densest rim of the RMC and are
presented by a subsequent paper (Li \& Smith 2004b, subm.). Regulated formation of clusters and
loose aggregates of embedded young stars toward the south-east boundary of the RMC
will be the subject of the third paper. This series, as a whole, 
aims at shedding
light on the long-puzzling issues related to sequential cluster formation in
galactic molecular clouds such as the RMC.
%what properties these compressed layers might have in producing new 
%generations of young stars, how stars and/or clusters are formed with 
%perhaps largely different but solid initial conditions and could this be taking place 
%in a much different manner?  Will it also follow a universal Initial Mass
%Function (IMF), shall they show up as bound clusters and what a role they 
%might play in the whole regime 
%of star formation? 
The exceptional geometry of the Rosette Nebula and its 
environment make this region an ideal target for this study. 
Our new information is derived from archived 2MASS data. In
contrast to visual inspection, our methodology permits us to
identify and analyze the fine structures of clustered star formation in the RMC.

%%%%%%%%%%%%%%%%%%%%%%%%%%%%%%%%%%%%%%%%%%%%%%%%%%%%%%%%%%%%%%%%%%%%%%%%
\section{Data acquisition \& sample selection}
%%%%%%%%%%%%%%%%%%%%%%%%%%%%%%%%%%%%%%%%%%%%%%%%%%%%%%%%%%%%%%%%%%%%%%%%
\label{data}

The 2MASS is the first near infrared project that
made uniformly-calibrated observations of the entire sky in the J (1.24 $\mu$m),
H(1.66 $\mu$m) and Ks(2.16 $\mu$m) bands with a pixel size of 2.0{\arcsec}. Sources
brighter that about 1 mJy in each band were detected with a signal to noise
ratio greater than 10, which leads to a photometric completeness to 15.9,
15.0 and 14.3 mag, respectively, for each band in unconfused regions. For detailed
information about the 2MASS survey and the 2MASS Point Source Catalogue, please
refer to the 2MASS Explanatory Supplement.

Archived data in the 2MASS Point Source Catalogue (PSC) and IRAS Sky Survey 
Atlases (ISSA) were retrieved via IRSA (http://irsa.ipac.caltech.edu/). 
The following sample selection criteria were employed to guarantee the reliability of
the 2MASS data in use and to allow a rigorous analysis. (1). Each
source extracted from the 2MASS PSC must have a certain detection in
the J, H \& Ks bands. This constrains the 'rd-flg' (data reduction
flag) to only 1, 2 or 3 in each digit and the 'ph-qual' (photometric
quality) to only A, B, C or D for each band. (2). We require the
Ks band signal to noise ratio, K-snr, to exceed 15. We find that
this further constrains field stars in the control field to the main-sequence 
and post-main sequence loci on the
(J-H) -- (H-Ks) diagram, as will be illustrated in Sect. 4.
However, it should be noted that the flux and resolution
limited 2MASS survey, and our stringent source selection in combination will result
in drawbacks.  Firstly, the number of young massive stars that are deeply embedded can be
underestimated; this is likely to occur in the densest regions with high extinction.
Secondly, this work is limited to the study of comparatively high mass cloud members.
With a 90$\%$ completeness limit of $\sim$~0.8~M$\odot$, even disk stars with masses
above this limit can be missed.

%%%%%%%%%%%%%%%%%%%%%%%%%%%%%%%%%%%%%%%%%%%%%%%%%%%%%%%%%%%%%%%%%%%%%%%%
\section{Methodology}
%%%%%%%%%%%%%%%%%%%%%%%%%%%%%%%%%%%%%%%%%%%%%%%%%%%%%%%%%%%%%%%%%%%%%%%%

Due to its spatial proximity to the Galactic plane and its location at a distance of 
$\sim$ 1.4 kpc, a vast number of foreground and background  field stars are expected 
in the infrared towards the RMC. Despite this heavy contamination, particularly by foreground
stars, two compact stellar clusters are immediately apparent from the spatial distribution 
of all the sources extracted from the 2MASS catalogue. One cluster is located right at the 
centre of the Rosette Nebula and thus corresponds to the young open cluster NGC\,2244. 
The other, situated at the densest rim of the RMC, has no optical counterpart 
and can well be a rich embedded cluster.  

\begin{figure*}
\centering
\includegraphics[width=8.7cm,height=6.35cm]{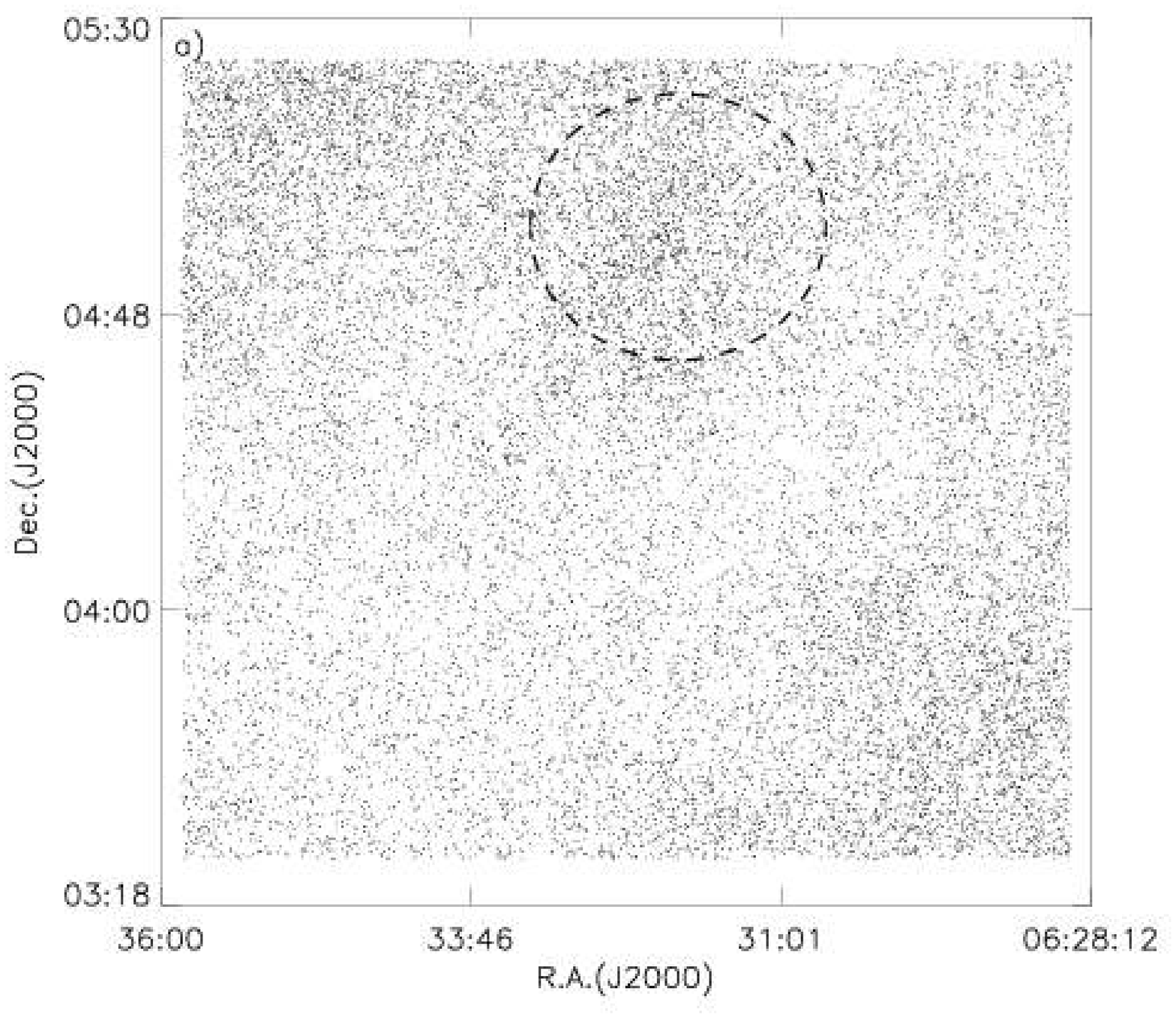}
\includegraphics[width=8.7cm,height=6.42cm]{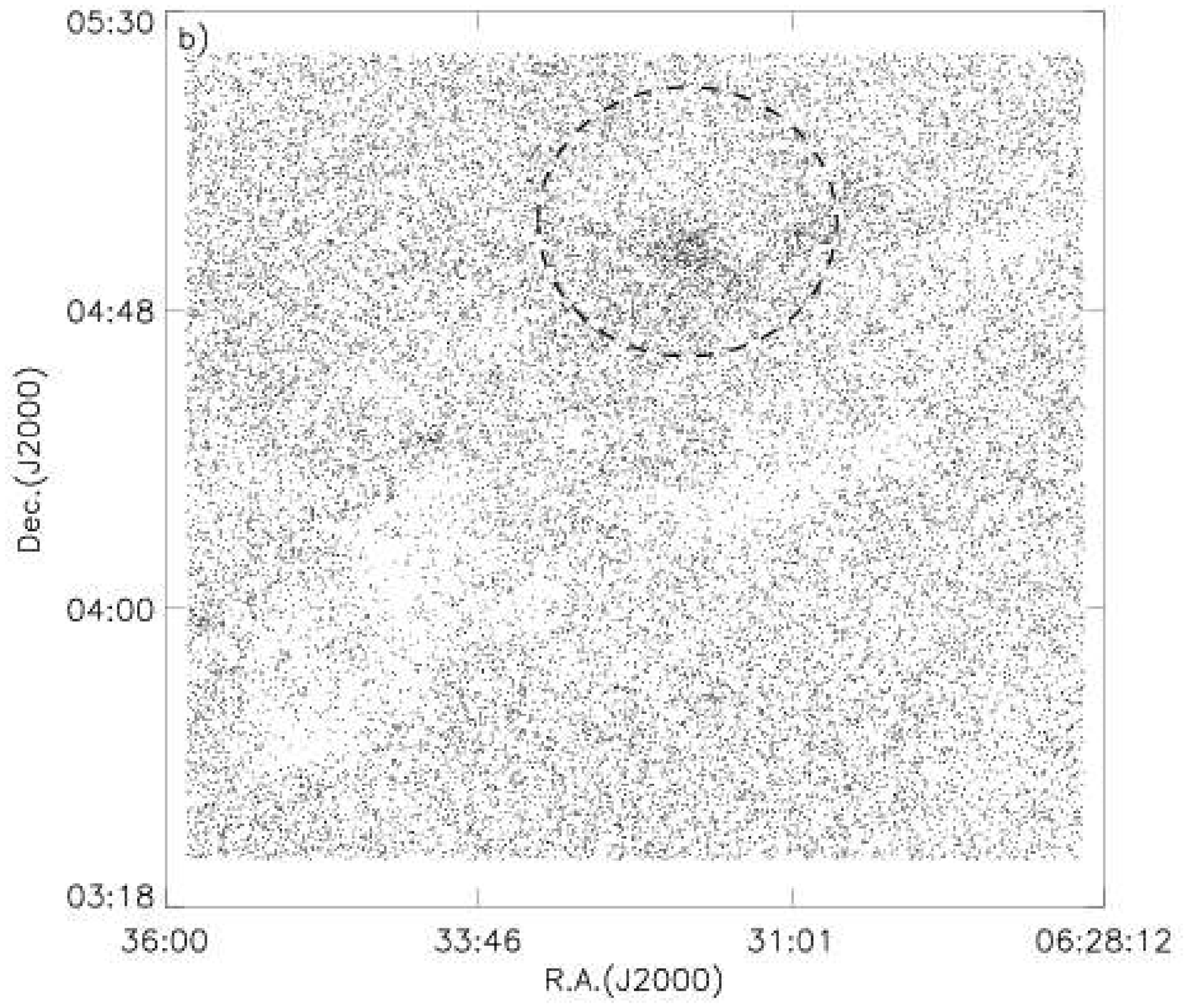}
\includegraphics[width=8.7cm,height=6.3cm]{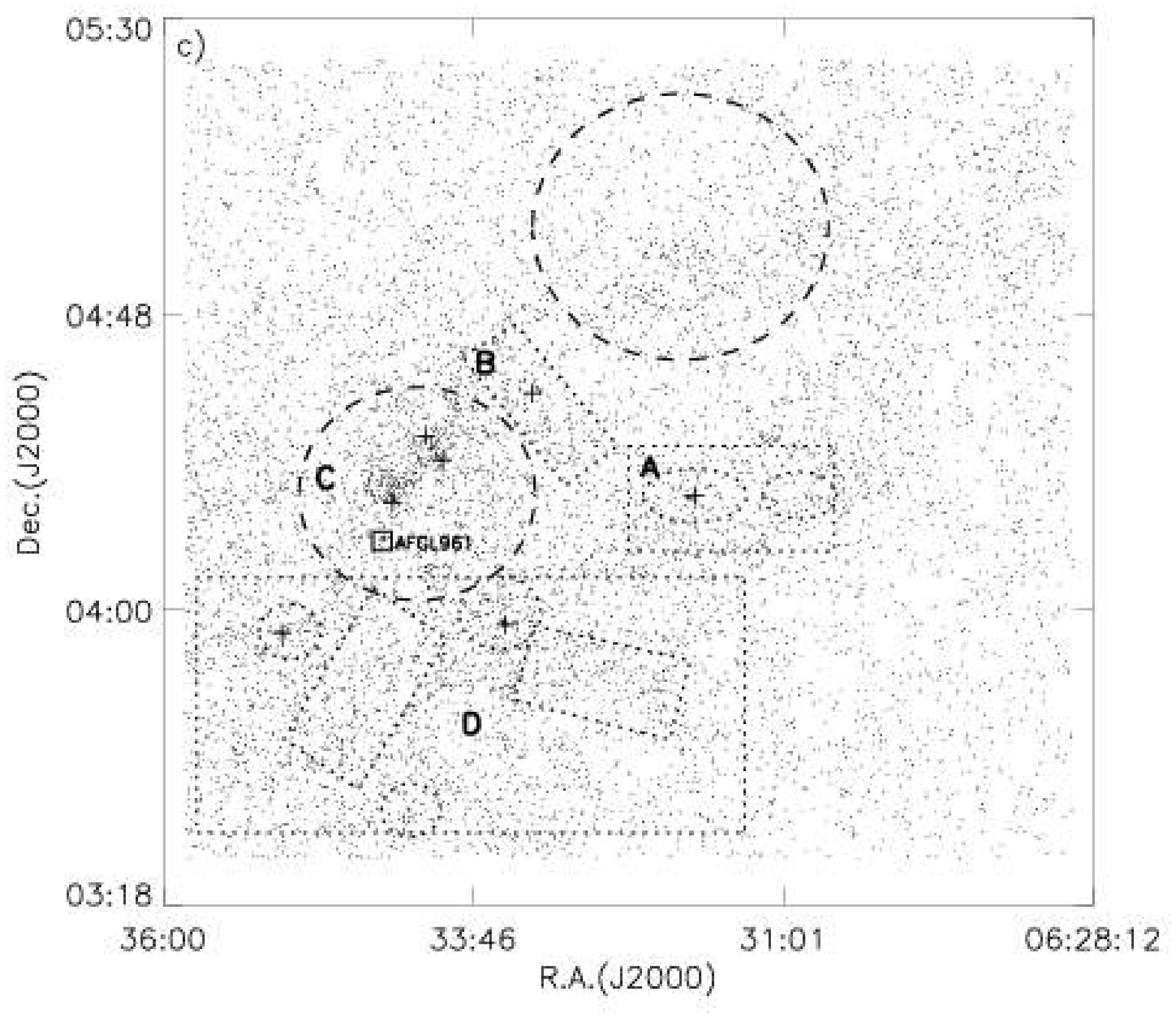}
\includegraphics[width=8.7cm,height=6.3cm]{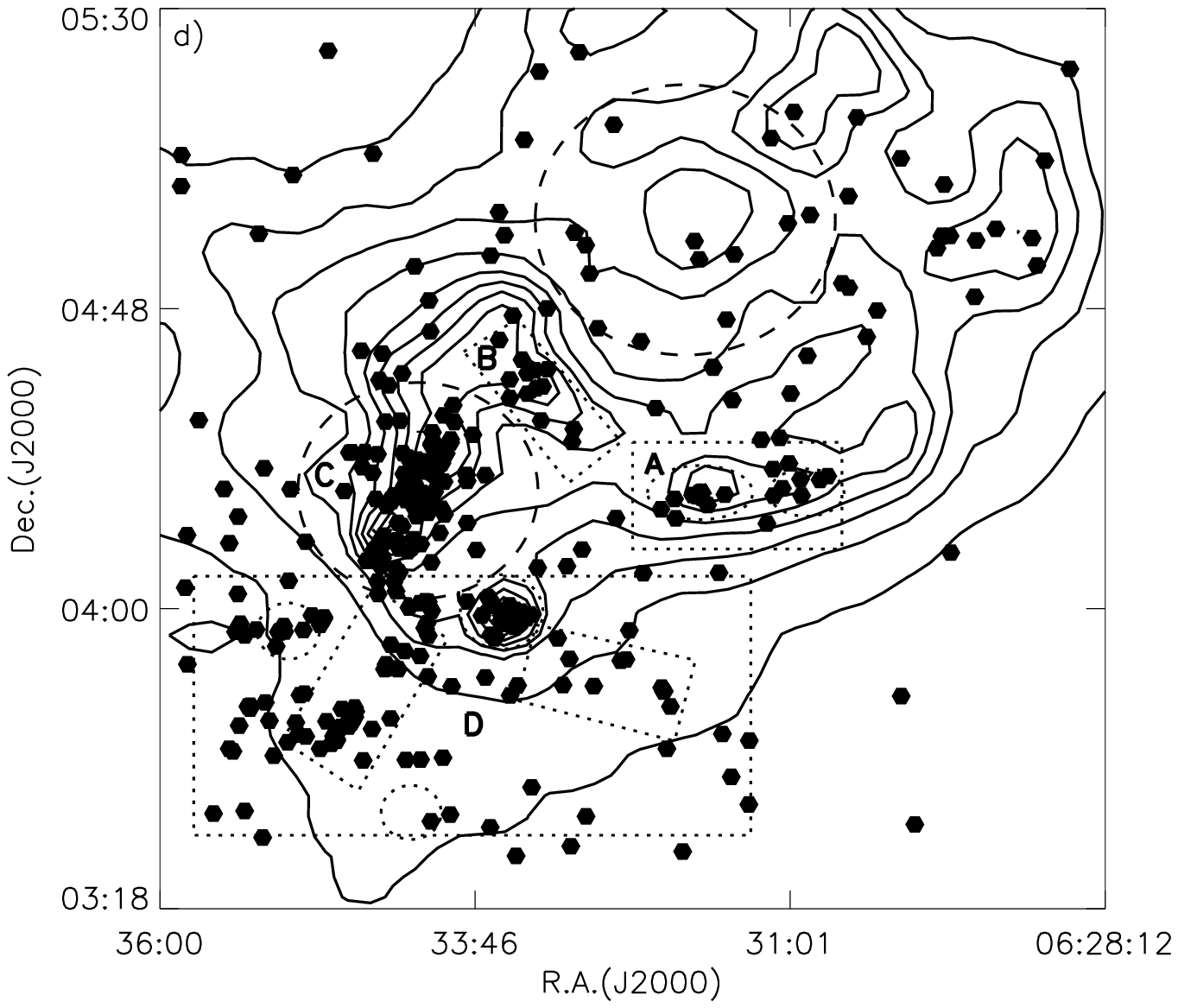}
\caption{The spatial distribution of the 2MASS sources toward the RMC as a function 
of the H--Ks colour: a) sources with H--Ks\,$<$\,0.2, b) 0.2\,$<$\,H--Ks\,$<$\,0.5,
c) H--Ks\,$>$\,0.5, d) H--Ks\,$>$\,1.0\,mag. Contours of the IRAS 100$\mu$m emission are
overplotted to elucidate the spatial distribution of the embedded clusters 
with respect to the highly clumped cloud structure. Well defined regions of 
clustered star formation are marked as A, B, C \& D with the 
extent of stellar groups further delineated by
dotted circles or rectangles according to their individual appearance.
The dashed circle drawn at the upper right of all plots indicates
the location of the young open cluster NGC\,2444 and, consequently, the 
H{\small II} region.}
\label{Fig1}
\end{figure*}

However, when the spatial distribution of the 2MASS sources is plotted as a function of the 
H--Ks colour, fine structure signifying large scale clustering is apparent all through the 
south-east part of the RMC (Fig.~\ref{Fig1}).  The 2MASS sources plotted in Fig.~\ref{Fig1}a 
are restricted to H--Ks $<$~0.2\,mag, a mean colour typical of main-sequence dwarves.  
A widely uniform distribution which contains the vast majority of foreground stars in 
this field is displayed.  The two dense regions in the north-east and south-west corners of 
the plot indicate field areas with very low extinction, where a combination of 
foreground and background sources is present. The low extinction regions are anticorrelated 
with regions of high CO density \citep{1995ApJ...451..252W} or dust opacity associated with the RMC. 

The distribution displayed in Fig.~\ref{Fig1}b with 0.2~$<$ H--Ks~$<$ 0.5, on the other hand, 
is equivalent to a local extinction map of the RMC. Foreground stars are excluded and 
a majority of the background field stars are obscured by the condensed regions with high extinction.
The higher the extinction, the lower the number density of the 2MASS sources. Note that 
members of the young open cluster, NGC\,2244, are prominent in this particular plot because of
the apparently low extinction along the line of sight to and through the blister 
H{\small II} region.  

The distributions of stars with higher H--Ks colour indices are shown in Fig.~\ref{Fig1}c 
\& \ref{Fig1}d in which minimum H--Ks values of 0.5 and 1.0, respectively, are taken. There is clear evidence for 
multi-seeded clustering at various places across the complex. This 
indicates that the RMC is a unique region for the study of sequential cluster formation,
especially when, for the first time, medium to low-mass stars are involved.

%%%%%%%%%%%%%%%%%%%%%%%%%%%%%%%%%%%%%%%%%%%%%%%%%%%%%%%%%%%%%%%%%%%%%%%%%%
\section{Clusters formed in the swept-up shell}
%%%%%%%%%%%%%%%%%%%%%%%%%%%%%%%%%%%%%%%%%%%%%%%%%%%%%%%%%%%%%%%%%%%%%%%%%%
\label{results}

A clear density enhancement of highly reddened stars in the fragmented
interaction layer between the H{\small II} region and the surrounding molecular
cloud is evident in Figs.~\ref{Fig1}c \& \ref{Fig1}d. Four or more compact far
infrared emission regions form a circular structure around the expanding
H{\small II} region  (Fig.~\ref{Fig1}d), powered by the strong stellar winds from the 
dozens of OB stars of the young open cluster NGC\,2244. However, significant clustering
of sources with line of sight extinction is the most prominent in the south and south-east 
arcs of the broken shell structure. These two regions are known to be associated with
prominent IRAS point sources (Cox et al. 1990) and recent star formation (Phelps \& Lada 1997),
indicating the existence of probably extensive young clusters.  Nevertheless, minor 
clustering of sources and likely the association with active star formation seems to 
present also in the north-west fragments of the shell structure. 

\subsection{Spatial distribution}

\begin{figure}
\centering
\includegraphics[width=8.4cm]{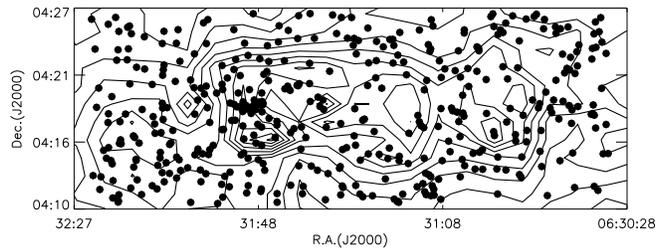}
\includegraphics[width=8.4cm]{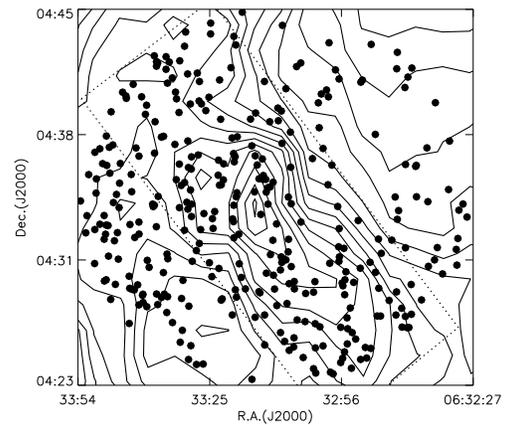}
\caption{Spatial distribution of the sources in Regions A (upper panel) \& B 
(lower panel). Subcluster
members are overplotted on the distribution of optical depth at 100\,$\mu$m of the 
corresponding fields, which clearly illustrate the well-confined shell structures.
The samples with H--Ks\,$>$\,0.5 are displayed.}
\label{ABdist0.5}
\end{figure}

The well-defined Regions A \& B are as outlined by dotted rectangles in 
Figs.~\ref{Fig1}c \& d. 
A close-up view of the clusters in Regions A \& B is presented in Fig.~\ref{ABdist0.5}. 
Sources with H--Ks~$>$~0.5\,mag are overplotted on the optical depth distribution of the 
compressed arcs. The optical depth at 100$\mu$m is derived from the IRAS ISSA data 
\citep[][and references therein]{1996ChA&A..20..445L}.

For region B, evidence for a cluster core is marginal (lower panel). Instead, loose 
aggregates of reddened young 
stars can be seen surrounding the central area where the highest optical depth is indicated.  
The distribution seems to follow well the compressed layers of the shell structure as 
presented by contours of the optical depth. However, a largely uniform distribution of 
group/cluster members is observed within the swept-up arc resembling a tilted rectangle, 
which is believed to be confined by the radiation pressure and impact from the stellar 
wind of the young OB cluster NGC 2244. 

A more or less similar stellar distribution is found in Region A, where two 
groups of reddened stars are identified, concentrated around two significant opacity cores, one 
on each side (upper panel of Fig.~\ref{ABdist0.5}). The seemingly
organized stellar distribution, following the contour lines of opacity, can also
be a reflection of the extinction effects on background field stars. As opacity
is indicative of the associated gas density, could the regulated distribution 
indicate an iso-density nature of the compressed layers of materials traced by the contours 
of optical depth in each of the fragmented arcs?
%There is a prominent concentration of sources in the left centre of the 
%region that associated with the location of the highest optical depth.

%-----------------------------------------------------------------------------

\begin{figure}
\centering
\includegraphics[width=8cm]{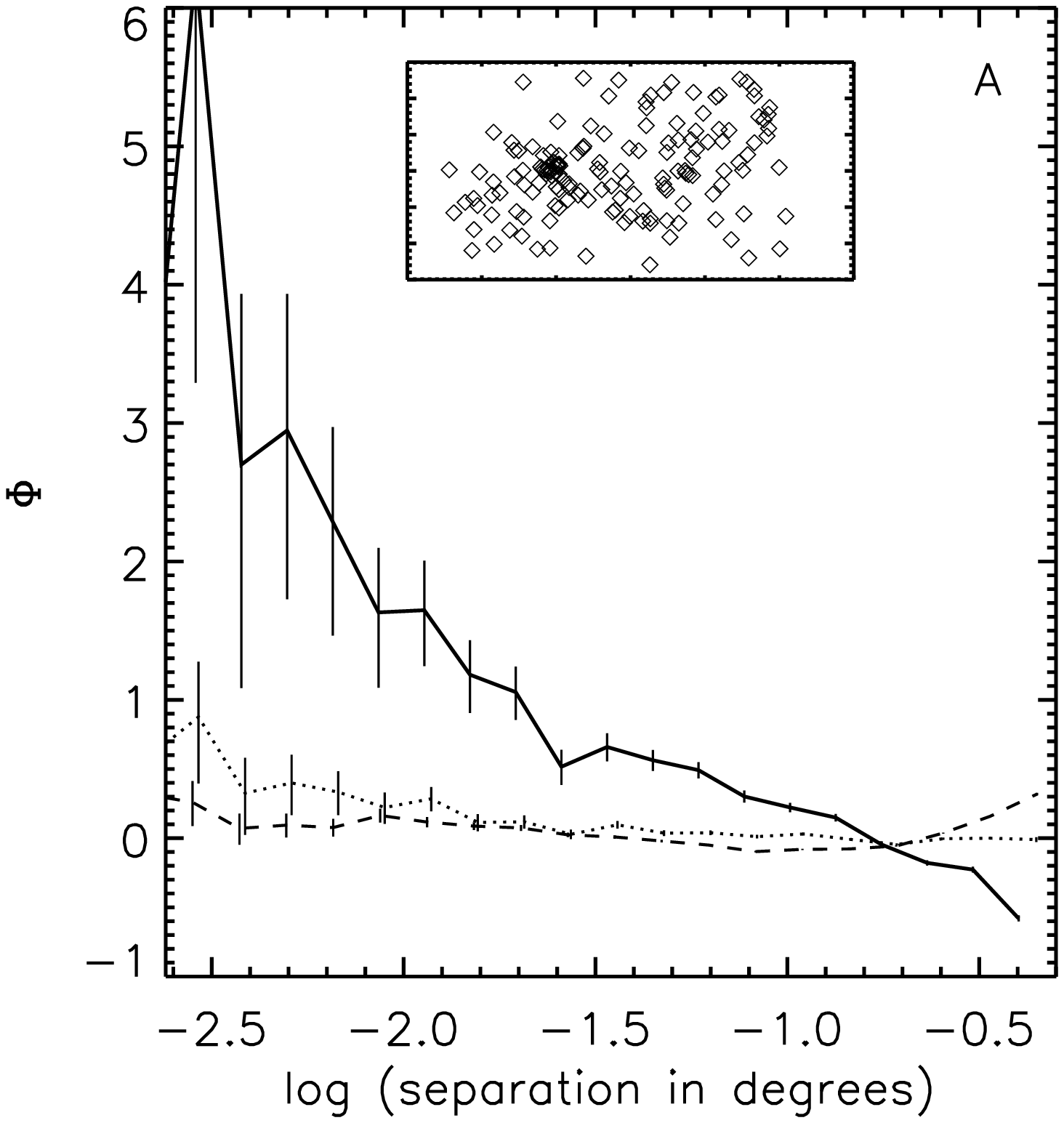}
\includegraphics[width=8cm]{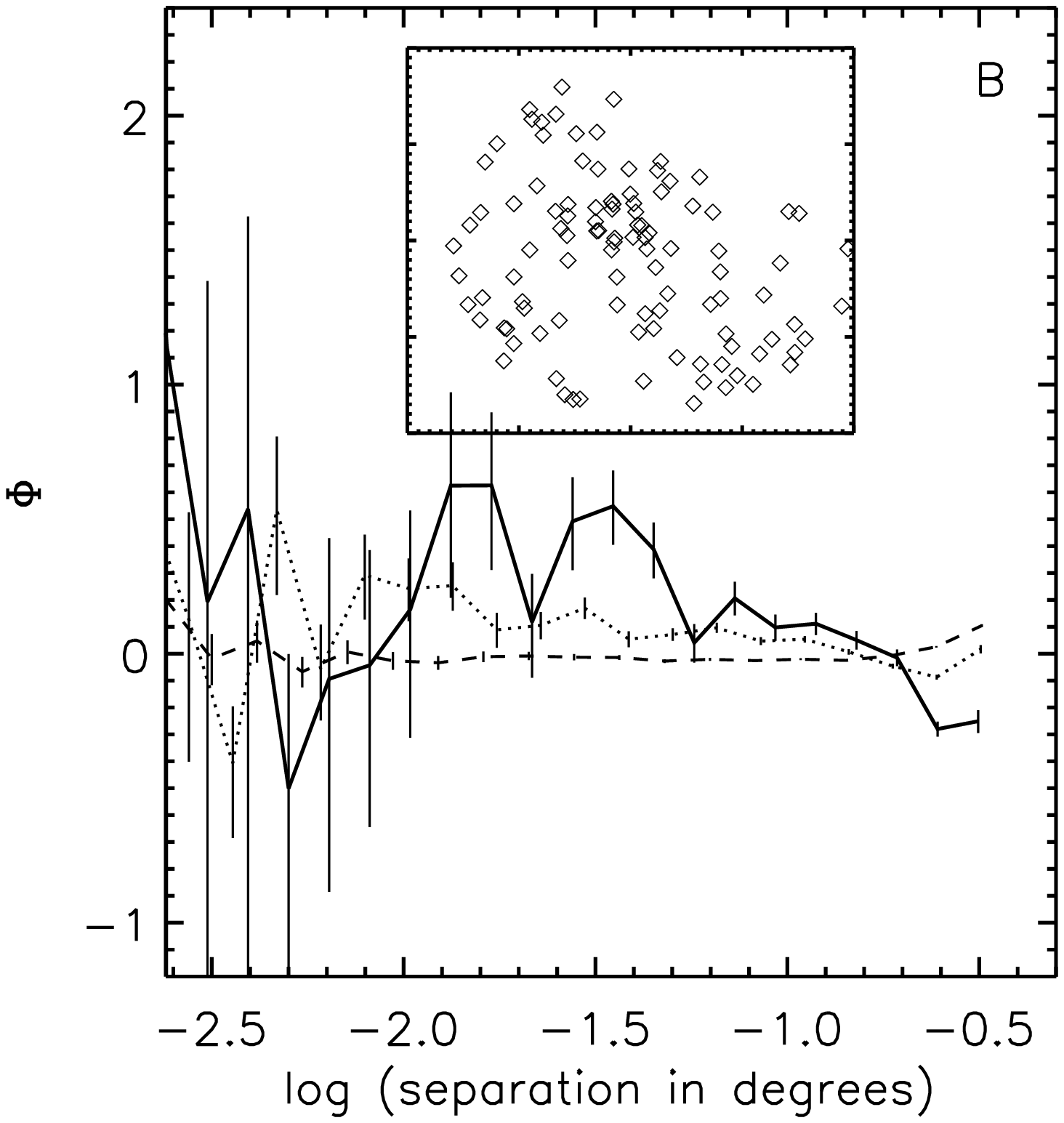}
\caption{The two-point correlation functions for the clusters embedded in Regions A 
(top panel) \& B (lower panel). The full, dotted and dashed lines correspond to
samples  with H--Ks $>$\,0.7,  $>$\,0.5 and  $>$\,0.2, respectively. The inset panels
display the spatial distributions for  H--Ks $>$\,0.7. The error bars correspond to
$\surd$N statistics.}
\label{twopoint}
\end{figure}

To check the significance of clustering in each target region, a two-point correlation function 
is applied and the results of the calculations are presented in Fig.~\ref{twopoint}. Within the 
selected areas, we calculate the number of pairs as a function of angular separation, r, as 
$H_d(r)$. This is normalised to the values expected from a randomly distributed sample
spread over the same area of the sky, $H_r(r)$. We plot the function 
\citep[e.g. see][]{2004SSGK}
\begin{equation}
\Phi = \frac{H_d(r)}{H_r(r)}  -   1,
\end{equation}
clustering is implied where $\Phi$ exceeds zero. 

We uncover hierarchical clustering in Region A but predominantly
for the sample of highly reddened sources. As displayed
in Fig.~\ref{twopoint}, the higher the H--Ks colour constraint imposed on the sample,
the clearer the correlation. The correlation function substantiates the clear 
concentration of sources in regions indicative of active star formation \citep{1997ApJ...477..176P}
which cannot result from a random distribution
of background field stars. On a log-log plot (not shown) the power law index
of $\Phi$(r) for Region A is -1.2, very steep in comparison to stellar clusters. This
suggests that the embedded clusters are very young and that no flattening has been caused 
by the stellar velocity dispersion.
% This, however, implies that we are quantifying the 
%spatial distribution of extinction rather than the intrinsic stellar distribution,
%which is probably better represented by the function for H--Ks\,$>$\,0.5.  As
%shown by the dotted line in Fig.~\ref{twopoint}, no clustering is
%found for these young stars even though they are also deeply embedded.

However, only minor signatures of clustering, on scales of order 0.1$^\circ$, are found 
in Region B (Fig.~\ref{twopoint}, lower panel) due to the loose distribution of the
candidate cluster members.  As noted in Sect. 1,
given an age of 1~Myr, a cluster might expand by $\sim$ 5\arcmin~ in 1\,Myr.
Therefore, even the absence of clustering is not evidence for non-clustered star 
formation assuming that all the reddened sources are embedded. Furthermore,
the clustering of sources in Region B is apparent on large scales (Fig. 1c \& 1d).

%------------------------------------------------------------
%------------------------------------------------------------

%The large-scale spatial distribution of the cluster members along the expanding H{\small I} shell 
%can be offering new insights into induced star formation in interaction layers. Massive clusters such as
%NGC 2244 may also have formed in this manner. The common shell properties imply a star
%formation scenario different from other processes such as isolated or clustering star formation 
%as a result of cloud fragmentation and collapse driven by turbulence or instability. 

%No substantial compact cores are detected in both Regions A \& B, though marginal cores
%do exist. This, along with the nearly uniform
%distribution of the cluster members as discussed above, favours
%a coeval or even simultaneous star formation scenario in what could be
%largely iso-density, iso-pressure layers. This may also be the cause 
%of the gravitationally unbound nature of the newly formed clusters,
%the inevitable fate of which is probably dynamical dispersion. They may finally end up 
%as field stars after the interstellar gas is evaporated on a comparatively short 
%time scale on the order of several Myrs. On the other hand, due to their
%origin as externally triggered, the cluster members are 
%exceptionally coeval and may be taken as important samples of destination
%oriented studies which demands more restrict constraints on evolutionary ages.

%, demanding a model with a strong restriction
%on age spread.
%Jinzeng: no idea what this means!! as samples in destination

%%%%%%%%%%%%%%%%%%%%%%%%%%%%%%%%%%%%%%%%%%%%%
\subsection{Colour-Colour Diagram}
%%%%%%%%%%%%%%%%%%%%%%%%%%%%%%%%%%%%%%%%%%%%%

\begin{figure*}
\centering
\includegraphics[width=8.7cm,bb=99 360 476 720]{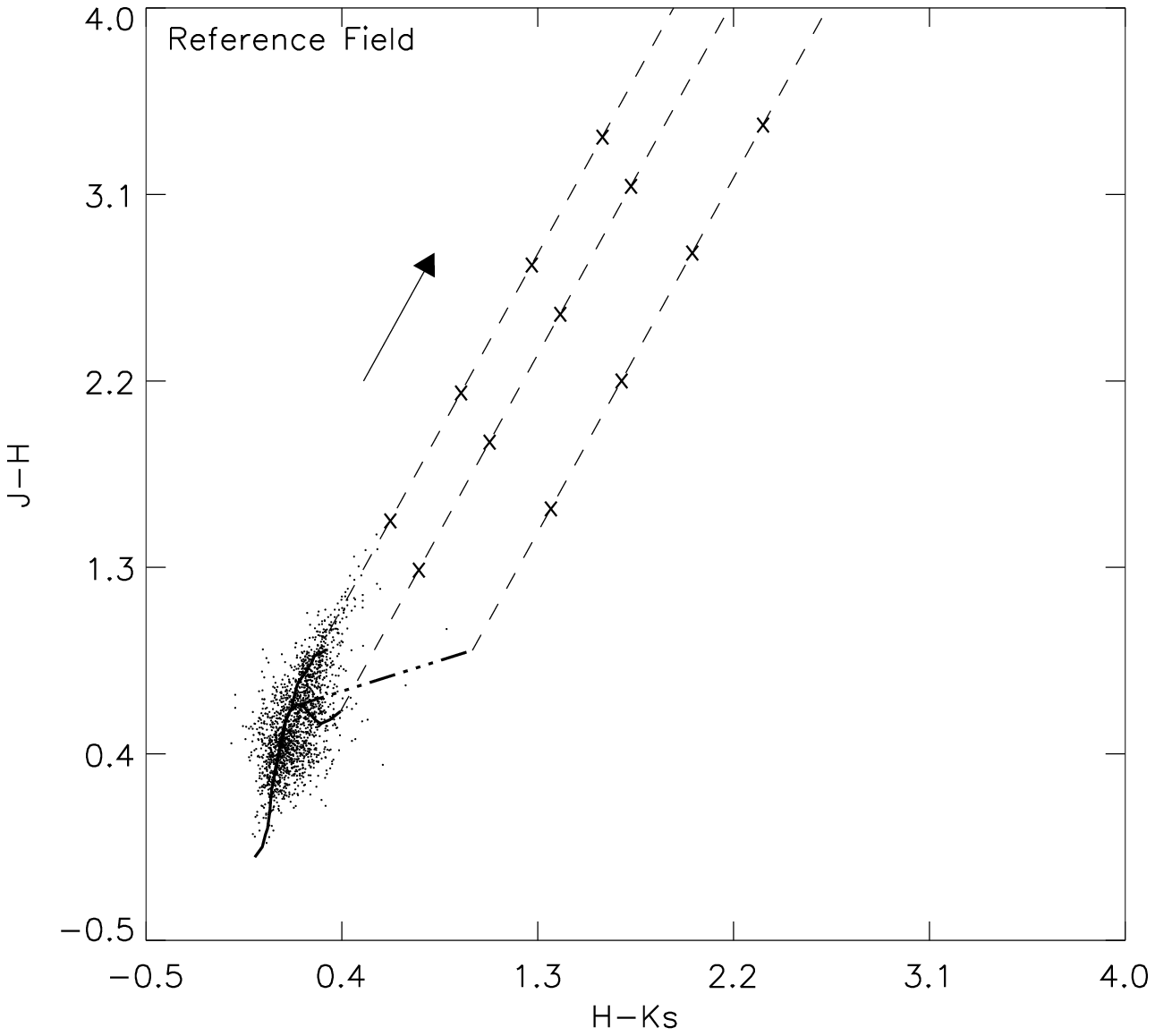}
\includegraphics[width=8.7cm,bb=99 360 476 720]{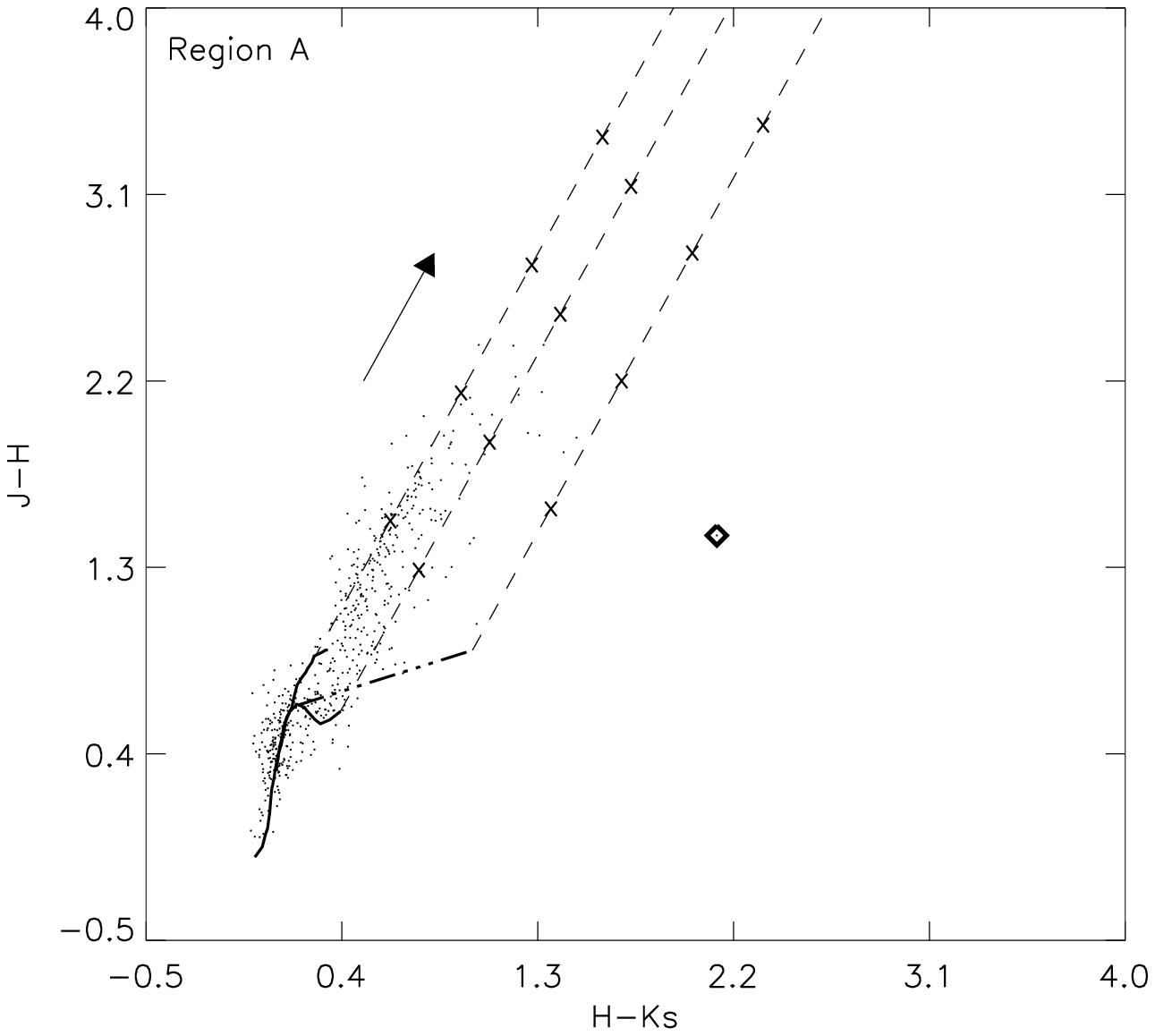}
\includegraphics[width=8.7cm,bb=99 360 476 720]{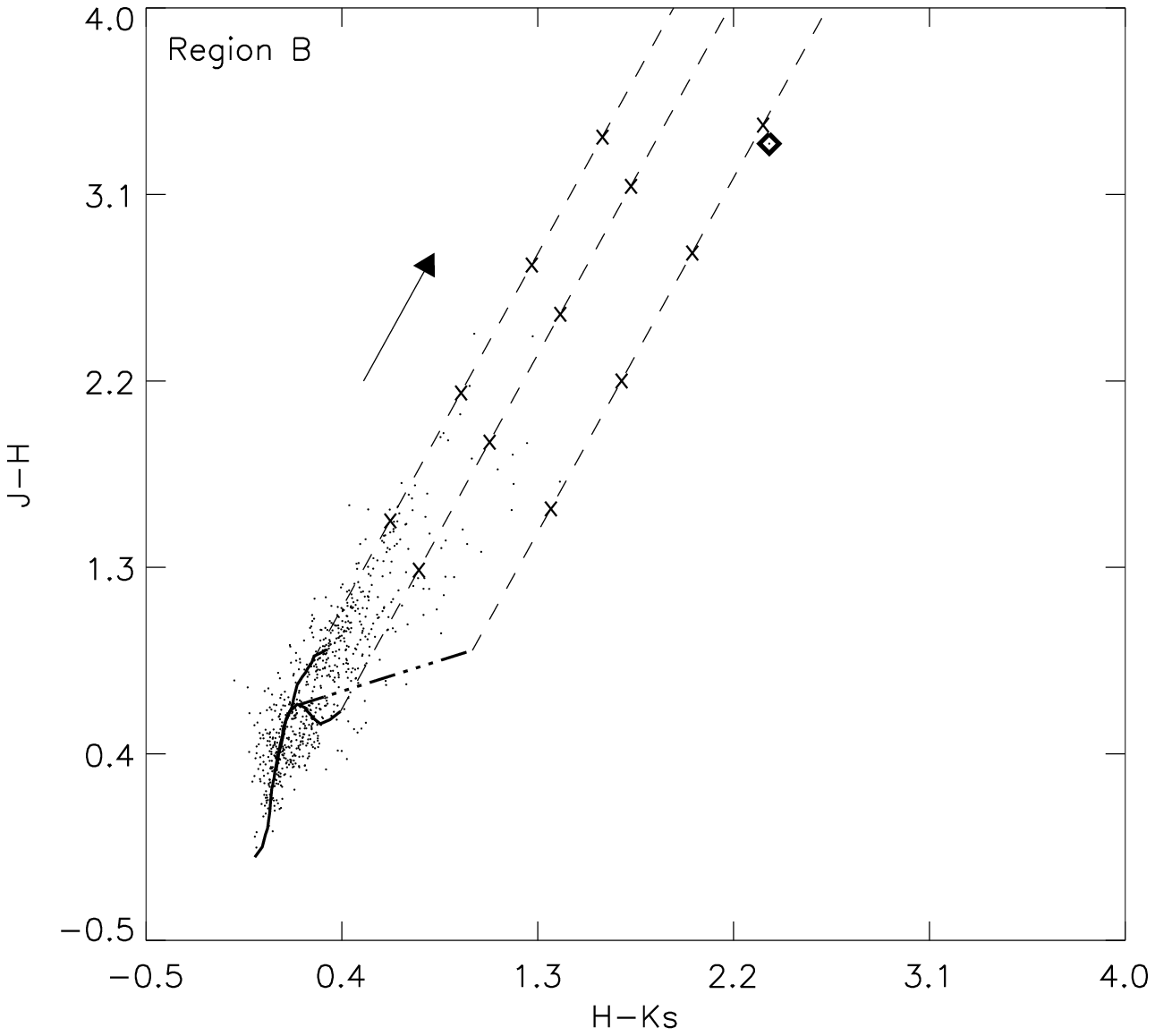}
\caption{Colour-Colour diagrams of (a) the reference field with a diameter of 40\arcmin and 
the clusters in (b) Region A \& (c) Region B. 
The source with the highest H--Ks colour excess in each cluster is overplotted with a bold 
type diamond. Solid lines represent the loci of the main-sequence dwarves and giant stars 
\citep{1988PASP..100.1134B}. The arrow in the upper left of the plot shows a reddening vector 
of A$_V$\,=\,5~mag \citep{1985AJ.....90..900R}. The dot-dashed line indicates the locus of
dereddened T~Tauri stars \citep{1997AJ....114..288M}. The dashed lines define the reddening 
band for the normal stars and T~Tauri stars, drawn parallel to the reddening vector. Crosses 
are overplotted at intervals corresponding to 5~mag of visual extinction.}
\label{ccddiagrams}
\end{figure*}

Colour-colour diagrams help in revealing the nature of reddened stars and in determining 
individual extinction. We have constructed colour-colour diagrams for two control fields 
(each with a radius of 20\arcmin), one located to the south-west and the other to the 
north-east of the Rosette Nebula, where visual extinction is very low as confirmed 
through the Digital Sky Survey images of the entire region. 
The results for the south-west field are presented in Fig.~\ref{ccddiagrams}a for a 
comparison to those for the embedded clusters.
We thus exclude any possibility of the existence of foreground extinction.
It is clear that all 2MASS sources in the control fields, selected using the criteria introduced
in Sect. 2, are concentrated to the main-sequence and giant-branch loci which,
as a whole, have a mean H--Ks colour of about 0.2~mag. This indicates that there is
negligible extinction toward the control field and presumably also towards the RMC.

Colour-colour diagrams for Regions A \& B are illustrated in Figs.~\ref{ccddiagrams}b \& c. 
Currently, no further effort was made to distinguish the two subclusters embedded in
Region A. At first glance, the distributions of foreground dwarves and candidate cluster 
members are apparently separate in both diagrams. However, concentrated to the left
edge of the reddening band, there seems to be a spread of background giants affected
by differential reddening effects in each region.

A close investigation of the (J-H) -- (H-Ks) diagrams indicates a visual extinction 
between 0.2 and 15.0~mag in Region A, and a weighted mean extinction
of 7.0\,$\pm$\,0.5~mag. The line of sight extinction of Region B lies between 0.5 and 
19.5~mag but shows a weighted mean of only 2.5\,$\pm$\,0.5 mag. Assuming that both
regions have the same average gas density, the small mean extinction of Region
B might indicate that its distribution along the line of sight is shallow. This is 
consistent with its swept-up origin.

%------------------------------------------------------------------

Around one seventh of the reddened sources toward Region A and a fifth of those in Region B are 
located to the right of the reddening band, indicative of intrinsic excessive emission in the
near infrared. Both regions have one extreme source 
located to the right of the reddening band of classical T~Tauri stars. These sources are 
good candidates of intermediate-mass young stars associated with circumstellar disks
or envelopes, as will be further illustrated in the following subsection.
%** the excess of red culster members is evident in the CCDs.

However, note that the location of some embedded sources in the reddening band does 
not necessarily mean they are aged \citep{2001ApJ...553L.153H}. Some can be young 
cluster members with little or no excessive emission i.e. weak-line T~Tauri stars and 
their massive analogues, which can be hard to detect in the near infrared. This
is likely to occur in high density embedded clusters, 
where the interactions between stars and internally driven turbulence are strong. 
The lack of excessive emission of the young stars is otherwise attributed to external 
UV dissipation and ionization of the associated disks by their massive 
neighbours, as indeed occurs in Regions A \& B because of their spatial proximity to
the young OB stars of NGC 2244 \citep{1998A&A...335.1049S}.
% In addition, the heavy extinction of the Rosette
%Nebula complex can undoubtedly inhibit the detection of many or most
%medium to low mass background field stars. Therefore, the majority of the 
%candidate cluster members presented in this study would probably be cloud members.

%%%%%%%%%%%%%%%%%%%%%%%%%%%%%%%%%%%%%%%%%%%%%%
\subsection{Colour-Magnitude Diagram}
%%%%%%%%%%%%%%%%%%%%%%%%%%%%%%%%%%%%%%%%%%%%%%

\begin{figure}
%\centering
\resizebox{\hsize}{!}{\includegraphics[bb = 100 360 475 697]{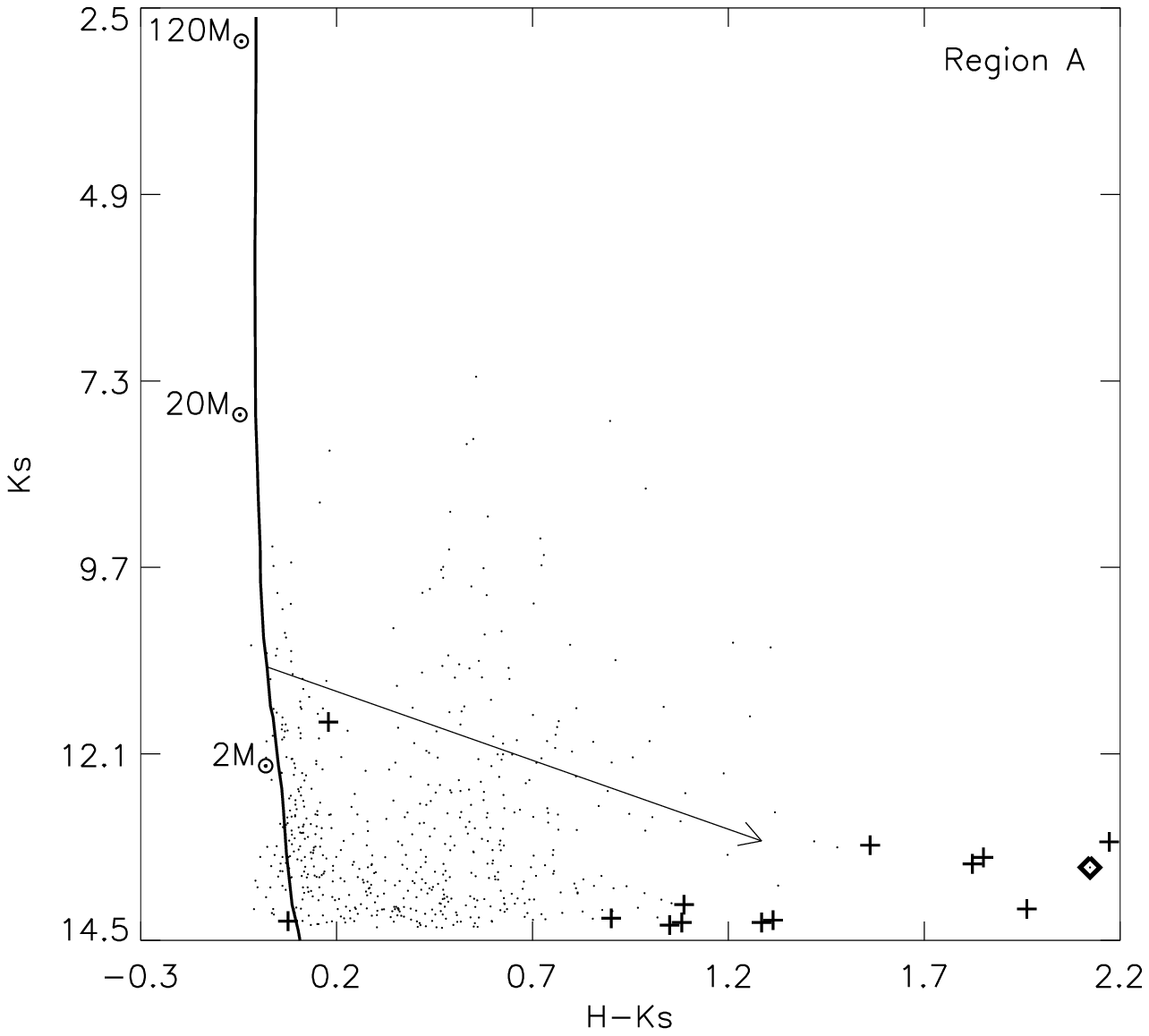}}
\resizebox{\hsize}{!}{\includegraphics[bb = 100 360 475 697]{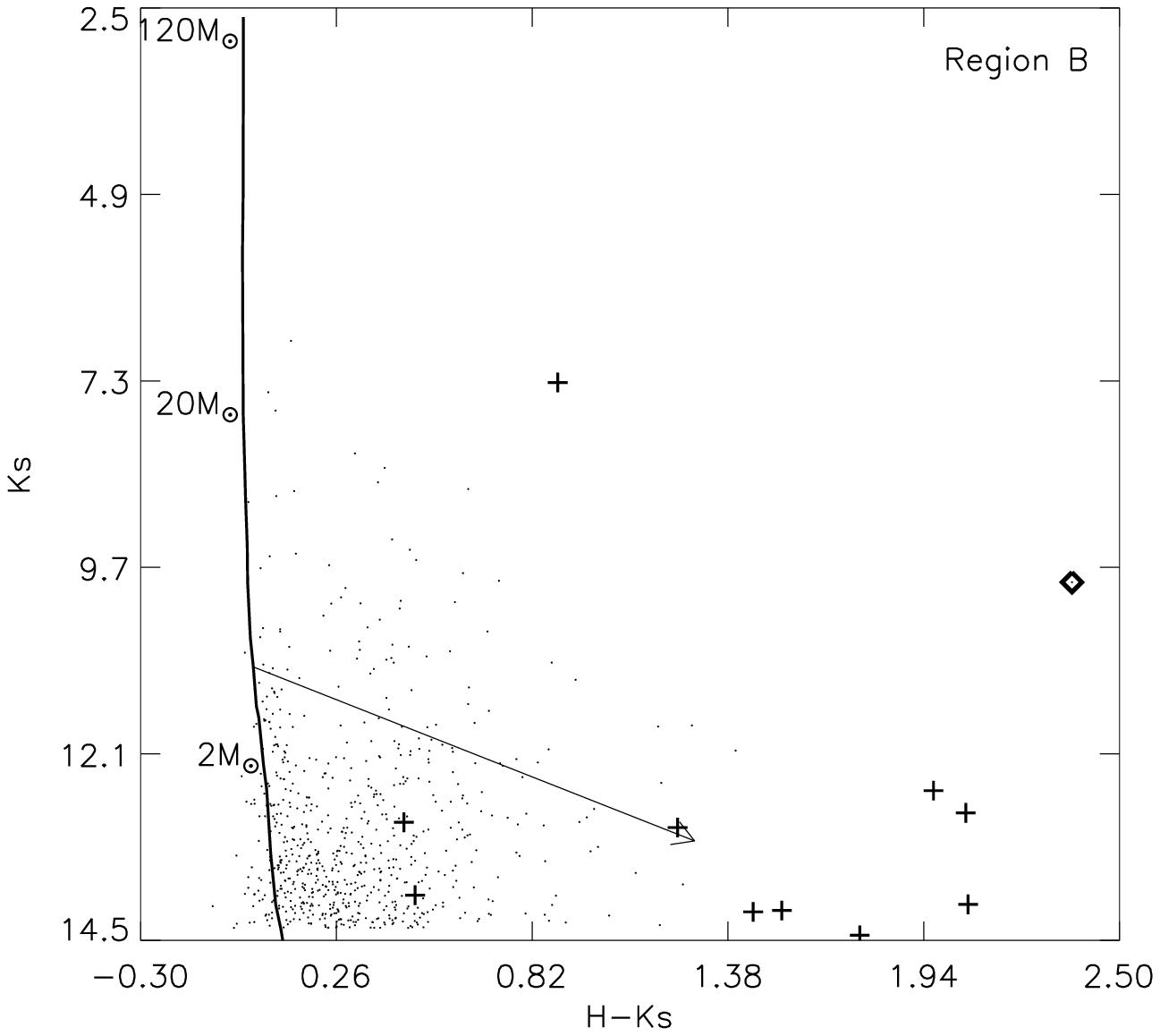}}
\caption{Colour-Magnitude Diagrams of the clusters in Regions A \& B. The source with the 
highest H--Ks colour excess in each diagram is
overplotted with a bold diamond, and criteria missed candidate young stars are 
presented as pluses. The main sequence for stars with masses between 0.8 and 120 M$\odot$
is plotted as a solid line \citep{2001A&A...366..538L}. The slanted line with an
arrow at the tip denotes a reddening of A$_V$ = 20~mag of a A0 type dwarf.}
\label{colormag}
\end{figure}

All 2MASS point sources that meet the criteria introduced in Sect. 2
are plotted on the Colour-Magnitude Diagrams (CMD) shown in Fig.~\ref{colormag}.
A clear separation between the foreground dwarves, which 
follow tightly the main-sequence presented by a solid line, and the reddened sources 
or candidate cluster members is apparent in the CMDs. The foreground main-sequence 
stars have a mean H--Ks colour of $\sim$\,0.2~mag, consistent with that determined 
from the colour-colour diagrams.  

More than one third of the candidate cluster 
members in Region~A and about one fourth of those in Region~B are located at positions
above the reddening vector drawn for an A0 dwarf. They are estimated to have masses
of below or around 20~M$_\odot$ and are therefore good candidates
for medium mass pre-main sequence stars still immersed in their natal molecular clumps.
This study based on the archived 2MASS data shows evidence of the formation of medium mass 
clusters in the fragmented interaction layer of the Rosette Nebula.
However, from the CMDs it can be deduced that many low-mass components of the
embedded clusters were missed.
New explorations with higher spatial resolution and sensitivity are needed to obtain 
a more complete census of the embedded population in these regions.
% as also indicated by the following subsection.

%As will be presented in the second of the series, massive compact subclusters containing 
%probably a proto-O stars group associated with AFGL 961 are being formed in Region C \citep{2004LSc} that
%situated to the south of the Rosette Nebula, where the visual extinction is 
%predominantly high based on CO \& C$^{13}$ detections of the RMC \citep{1980ApJ...241..676B,
%1995ApJ...451..252W}.  In Regions A \& B, 
%embedded stars with masses of only up to around 20~M$_\odot$ are nurtured. 
%Could the existence of an upper mass limit be a consequence of their
%formation in compressed layers? Here,
%all external and internal conditions responsible for star formation are set 
%rigid. A general high pressure may lower the Jeans mass throughout the shell.
%Or, such interaction layers may breed only medium to low mass stars simply because
%the near-simultaneous star formation activity provides a time scale too short for
%the development of higher mass molecular cores and consequently high-mass stars. 
%These star forming regions are thus exceptional samples for theoretical studies
%of induced clustering star formation, where the conditions are predictable.

%%%%%%%%%%%%%%%%%%%%%%%%%%%%%%%%%%%%%%%%%%%%%%%
\subsection{Missed candidates of young stellar objects}
%%%%%%%%%%%%%%%%%%%%%%%%%%%%%%%%%%%%%%%%%%%%%%%

The sample selection criteria employed in this study, along with the flux and 
resolution limited 2MASS survey, results in the exclusion of a number of stars 
with uncertain J band detection since it is more sensitive to extinction 
\citep{2004A&A...421..623K}, but high enough H and Ks signal-to-noise ratios.
Many of them were likely shrouded by bright sources either through 
projection or physical association, or rather severely affected by 
extensive emission in association or in the background.
These objects are mentioned here because they are predominantly sources with a 
highly reddened color and likely excessive emission.

This problem can be serious in the study of embedded clusters with high density 
or suffering from heavy differential extinction. This is found to be the case
from a detailed investigation of the compact subclusters in Region~C 
of the RMC \citep{2004LSc}, where a significant fraction of the protostellar 
candidates, including the most massive protostar AFGL 961E,
have been omitted from the data extracted from 2MASS. However, only a few 
disk star candidates are found to have been missed in Regions A 
\& B where both the stellar density and differential extinction are
comparatively low (Fig. 2). These objects are tentatively put on 
the CMD as an indication of the population of mostly candidate members
that did not meet the selection criteria (Fig.~\ref{colormag}).

%%%%%%%%%%%%%%%%%%%%%%%%%%%%%%%%%%%%%%%
\subsection{Ks Luminosity Function}
%%%%%%%%%%%%%%%%%%%%%%%%%%%%%%%%%%%%%%%

\begin{figure}
\centering
\resizebox{\hsize}{!}{\includegraphics[bb = 100 360 478 697]{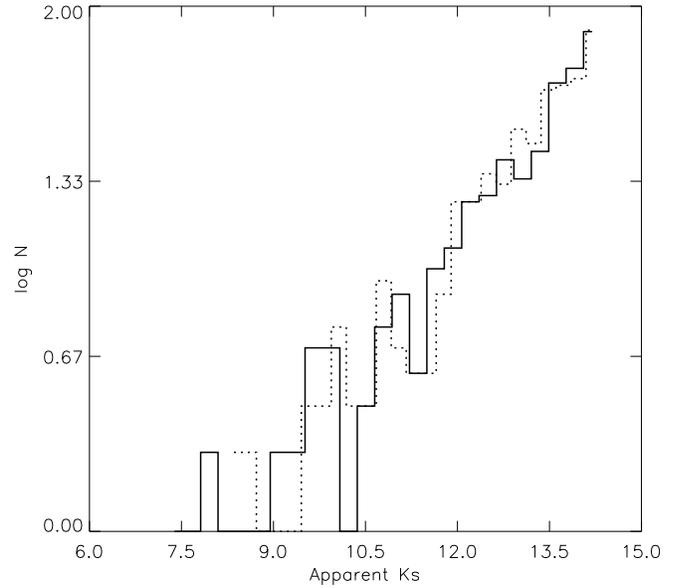}}
\caption{The Ks Luminosity Function of the clusters in Regions A \& B. The
KLF of the Region A cluster is plotted as a solid line, and of the Region
B cluster as a dotted line. It is evident that the clusters have
a very similar slope of ${>}$\,0.3 in the range 11.0\,${<}$\,Ks\,${<}$\,14.5,
signifying that they have ages of around 1~Myr and are still in their 
early stages of evolution.}
\label{lumfunction}
\end{figure}

The luminosity function of an embedded cluster can be taken as a direct
measure of the initial luminosity function of the cluster.
It is usually a statistically reliable tool for addressing questions
concerning the initial mass function, star formation history and pre-main sequence
tracks. The Ks band luminosity functions for Regions A \& B
are presented in Fig.~\ref{lumfunction}, with that of the Region A cluster indicated 
by a solid line and of the Region B cluster by a dotted line. We found
that the Ks luminosity functions of the two clusters are
identical, and follow a power law distribution. A linear least-square fit to the distribution in
the range 11.0\,${<}$\,Ks\,${<}$\,14.5 indicates a very similar slope for both
clusters of ${>}$\,0.3. This implies an age of $\sim$\,1~Myr for both the
Regions A \& B clusters from a comparison with other embedded clusters
with known ages \citep{1995AJ....109.1682L}. On the other hand, the Ks luminosity
function does not turn over at the 2MASS completeness limit, which in this study
corresponds to Ks\,=\,14.5~mag and $\sim$0.8~M$_\odot$ at a distance of 1.4~kpc. 
They are, therefore, still in a very early stage of evolution, within a few
arcminutes of their birth sites.

%-----------------------------------------------------------

Note that stars formed in such compressed shells seem to show a 
moderate upper mass limit, which could be a consequence of the existence
of a universal IMF.

%%%%%%%%%%%%%%%%%%%%%%%%%%%%%%%%%%%%%%%%%%%%%%%%%%%%%%%%%%%%%%%%%%%%%%%%%%%
\section{Summary \& Discussions}
%%%%%%%%%%%%%%%%%%%%%%%%%%%%%%%%%%%%%%%%%%%%%%%%%%%%%%%%%%%%%%%%%%%%%%%%%%%

Two medium-mass infrared clusters, associated with the expanding
shell driven by the Rosette Nebula, are examined in this paper. One is associated with Region
B of the fragmented shell and the other with Region A. These shell clusters are found to have an age of around 1~Myr, 
much younger than the young open cluster NGC\,2244 that excites the H{\small II} 
region as well as the dynamical age of the associated H{\small I} shell.

Only marginal evidence is found for cores in the embedded clusters. The
majority of the candidate cluster members are widely distributed along the
well-defined shell structures.  They present a largely uniform distribution
which could imply a gravitationally unbound future of the clusters when the 
associated gas is ultimately dissipated or exhausted. 
This case study may also provide insight into how large numbers 
of embedded clusters fall apart upon emergence (Lada \& Lada 2003).\\

The interaction layers between blister H{\small II} regions and their dense 
molecular surroundings, such as Regions A \& B of the RMC, turn out to 
be unique sites of star or cluster formation. These working surface layers may play 
an important role in star formation in that they shield 
the ambient molecular cloud from being disturbed 
by the expanding H{\small II} region and the immersed young open cluster such as 
NGC\,2244. Indeed, UV dissipation and ionization is found to 
scale logarithmically with the decreasing UV flux deep in the molecular
clouds along the radial direction of NGC 2244 \citep{1998A&A...335.1049S}.
This may contribute to, or in the vicinity
of the interaction layer play a major role in the collapse of the clumps and consequently 
in the formation of new stars or clusters. This issue was further elaborated in \citet{2004LSa}.

\begin{acknowledgements}

We are grateful to an anonymous referee for the constructive comments and 
suggestions made for this paper.
Beside the 2MASS Archive, this work also made use of the IRAS PSC \& ISSA data.
DSS Survey data (STSCI, funded by NSF) were also employed. Financial support
was provided by PPARC and SRF for ROCS, SEM.

\end{acknowledgements}

%%%%%%%%%%%%%%%%%%%%%%%%%%%%%%%%%%%%%%%%%%%%%%%%%%%%%%%%%%%%%%%%%%%%%%%%%%%%%%%%
\bibliographystyle{aa}
\bibliography{rosette}

\end{document}